\documentclass[aps, twocolumn]{revtex4}
\usepackage{amsmath}
\usepackage{graphicx}
\usepackage{bm}
\usepackage{amssymb} 
\usepackage{amsopn} 
\DeclareMathOperator{\trace}{tr}
\newcommand* {\vek}[1]{{\ensuremath{\bm{\mathrm{#1}}}}}
\newcommand* {\vekc}[1]{{\ensuremath{\bm{\mathcal{#1}}}}}
\newcommand* {\kk}{\vek{k}}
\newcommand* {\Ee}{\mathcal{E}}
\newcommand* {\fj}{\mathfrak{j}}
\newcommand* {\fm}{\mathfrak{m}}
\newcommand* {\Ds}{\displaystyle}
\newcommand* {\Ts}{\textstyle}
\newcommand* {\frack}[2]{{\Ts\frac{#1}{#2}}}
\newcommand* {\bohrmag}{\mu_\mathrm{B}}
\newcommand {\td} [2] {\frac{d #1}{d #2}}
\newcommand {\pd} [2] {\frac{\partial #1}{\partial #2}}

\begin{document}

\title{Spin Orientation of Holes in Quantum Wells}
\author{R Winkler$^{1,2}$, Dimitrie Culcer$^{1,2}$,
        S. J. Papadakis$^3$%
        \footnote{Present address: Johns Hopkins University, Applied
        Physics Laboratory, Laurel, Maryland 20723, USA}%
        , B. Habib$^3$ and M. Shayegan$^3$}
\affiliation{$^1$ Advanced Photon Source, Argonne National Laboratory,
Argonne, IL 60439, USA} 
\affiliation{$^2$ Northern Illinois University, De Kalb, IL 60115, USA}
\affiliation{$^3$ Department of Electrical Engineering, Princeton
University, Princeton, NJ 08544, USA}

\begin{abstract}
  This article reviews the spin orientation of spin-3/2 holes in
  quantum wells. We discuss the Zeeman and Rashba spin splitting in
  hole systems that are qualitatively different from their
  counterparts in electron systems. We show how a systematic
  understanding of the unusual spin-dependent phenomena in hole
  systems can be gained using a multipole expansion of the spin
  density matrix. As an example we discuss spin precession in hole
  systems that can give rise to an alternating spin polarization.
  Finally we discuss the qualitatively different regimes of hole
  spin polarization decay in clean and dirty samples.
\end{abstract}
\date{\today}
\maketitle

\section{Introduction}
\label{sec:intro}

Spin electronics is a quickly developing research area that has
yielded considerable new physics and the promise of novel
applications \cite{wol01}. The electrons in the conduction band of
common semiconductors like GaAs are characterized by a spin-1/2.
Holes in the topmost valence band, on the other hand, have an
effective spin-3/2 (Ref.~\cite{lut56}) which gives rise to many
novel features that are not present in the conceptually simpler case
of spin-1/2 electron systems.

In this article we review some of the intriguing phenomena related
with the spin orientation in spin-3/2 hole systems. We begin in
Sec.~\ref{sec:luttham} with a brief review of the Luttinger
Hamiltonian which forms the foundation for a theoretical description
of spin-3/2 hole systems in cubic semiconductors like GaAs. In
Sec.~\ref{sec:zeeman} we discuss the anisotropic Zeeman splitting of
two-dimensional (2D) hole systems. While a magnetic field $B$
perpendicular to the 2D plane gives rise to a large Zeeman
splitting, the splitting in an in-plane magnetic field is greatly
suppressed. However, it can be tuned, e.g., by varying the thickness
of the quasi-2D system by means of external gates. At $B=0$, the
Rashba spin-orbit coupling in an inversion-asymmetric 2D system is
characterized by an effective magnetic field oriented in the 2D
plane. As discussed in Sec.~\ref{sec:rashba}, the resulting spin
splitting in 2D hole systems behaves thus very similar to the Zeeman
splitting in an external magnetic field. In particular, Rashba spin
splitting can likewise be tuned by varying the thickness of the
quasi-2D system.

In Sec.~\ref{sec:muexp} we review the multipole expansion of the
spin density matrix that provides a more systematic understanding
of the unusual spin-dependent phenomena in hole systems. As a first
application of this general approach, we discuss in
Sec.~\ref{sec:muma} the multipole moments induced in a 2D hole
system by an in-plane magnetic field. Next we use the multipole
expansion to discuss in Sec.~\ref{sec:hprec} the spin precession in
2D hole systems which turns out to be qualitatively different from
the more familiar case of spin precession in spin-3/2 hole systems.
For example, the hole spin polarization and the higher-order
multipoles can precess due to the spin-orbit coupling in the valence
band, yet in the absence of external or effective magnetic fields.
Finally, we discuss in Sec.~\ref{sec:relax} the spin polarization
decay in hole systems. Here, an important parameter is the product
of the precession frequency $\Omega$ times the momentum relaxation
time $\tau_p$. Qualitatively different regimes can be distinguished
for dirty samples with $\Omega\tau_p \ll 1$, weak-scattering samples
with $\Omega\tau_p \gg 1$ and ballistic systems with $\tau_p
\rightarrow \infty$. In Sec.~\ref{sec:conclusion} we summarize our
results.

\section{Spin-3/2 Hole Systems: the Luttinger Hamiltonian}
\label{sec:luttham}

\begin{figure}[tbp]
 \centerline{\includegraphics[width=85mm]{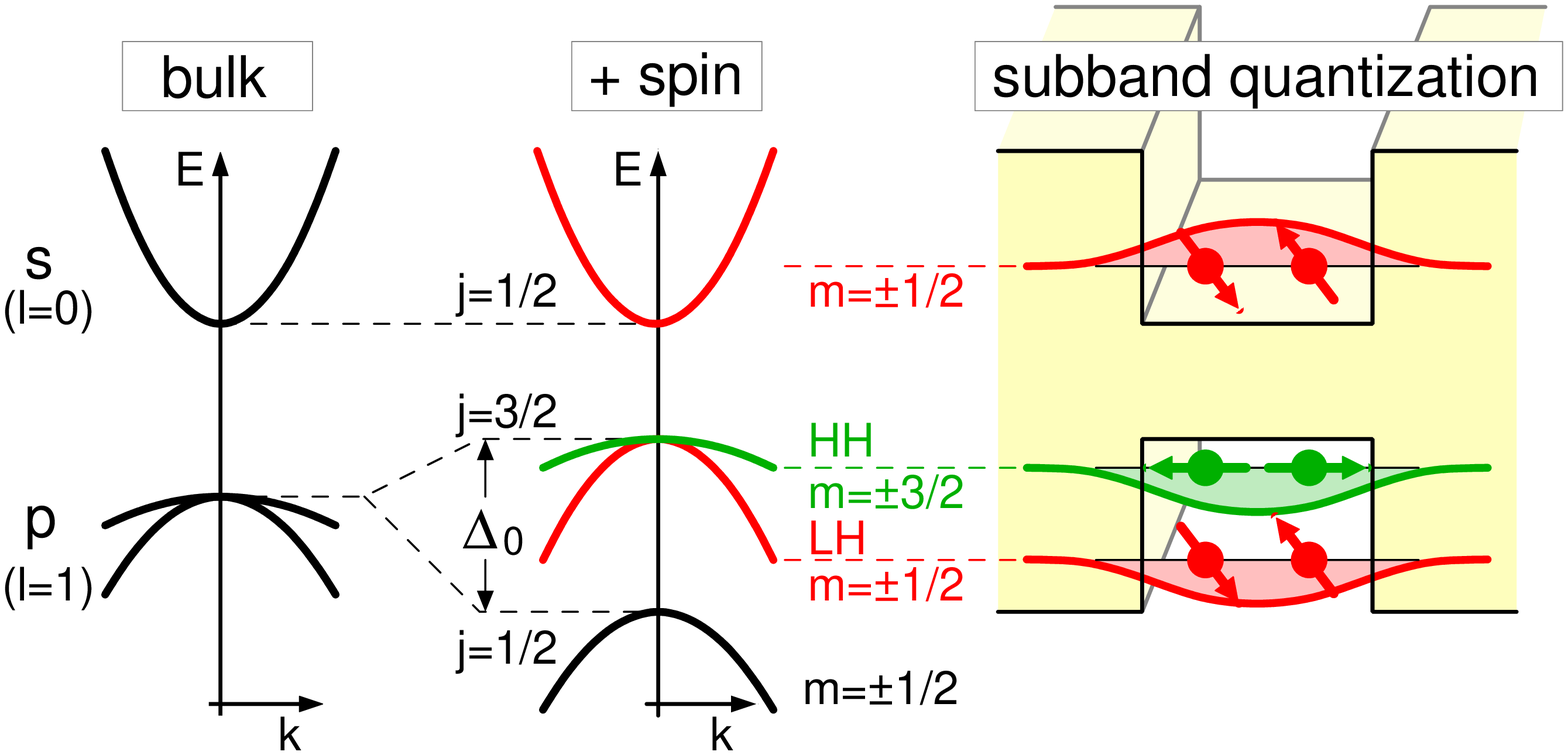}}
 \caption{\label{fig:hspin}Qualitative sketch of the band structure
 of GaAs close to the fundamental gap. (Left) The electrons in the
 conduction band are described by $s$-like atomic orbitals whereas
 the holes in the the topmost valence band are described by $p$-like
 atomic orbitals. (Center) Taking into account spin, the conduction
 band states are characterized by a total angular moment $j=1/2$,
 whereas in the valence band we have $j=3/2$ and $j=1/2$ states
 seperated by a spin-orbit gap $\Delta_0$. (Right) In quasi-2D
 systems the $j=3/2$ states split into HH and LH states. For
 electron and LH states the spin is essentially decoupled from the
 orbital motion, whereas HH states are predominantly made of $m=\pm
 3/2$ states with a spin quantization axis perpendicular to the 2D
 plane.}
\end{figure}

In a tight-binding picture, the electrons in the conduction band of
common semiconductors like GaAs are described by $s$-like atomic
orbitals \cite{yu01, win03}, see Fig.~\ref{fig:hspin} (left). Taking
into account spin, the electrons have a total angular momentum
$j=1/2$ that behaves analogously to a spin $s=1/2$. Holes, on the
other hand, are described by $p$-like atomic orbitals. Taking into
account spin, the holes have a total angular momentum $j=3/2$ and
$j=1/2$. The atomic spin-orbit coupling separates the $j=3/2$ from
the $j=1/2$ states by a spin-orbit gap $\Delta_0$, see
Fig.~\ref{fig:hspin} (center), so that we can associate an effective
spin $j=3/2$ with the states in the topmost valence band. At nonzero
wave vectors $k$ the fourfold degenerate states $j=3/2$ split
further into so-called heavy-hole (HH) and light-hole (LH) states.
If we choose the quantization axis of angular momentum $\vek{j}$
parallel to $\kk$, the HH (LH) states corresponds to $m=\pm 3/2$
($m=\pm 1/2$). (We ignore here small $k$-linear terms that couple HH
and LH states \cite{lut56}.)

In the following we consider quasi two-dimensional (quasi-2D)
systems in the $xy$ plane. Subband quantization in quasi-2D systems
correspond to standing waves in the $z$ direction so that we get a
splitting of HH and LH states called the \emph{HH-LH splitting},
even for in-plane wave vector $k_\|=0$, see Fig.~\ref{fig:hspin}
(right). For 2D systems grown on a high-symmetry (001) or (111)
surface, the $k_\|=0$ eigenstates are pure HH or LH eigenstates (if
we neglect the small $k$-linear terms). For any other surface, e.g.,
the experimentally important (113) surface \cite{dav91, her94}, we
obtain a weak coupling of HH and LH states even at $k_\|=0$ which is
caused by the terms of cubic symmetry in the Hamiltonian (see
below).

For $k_\| \ne 0$ we get a mixing of HH and LH states. However, for
typical sample parameters (well width and Fermi wave vector $k_F$),
the wave vector $k_z$ characterizing the standing wave perpendicular
to the 2D plane is generally much larger than $k_F$, $k_z \gg k_F$,
so that we can interpret the subband states even for $k_\| \ne 0$ as
HH or LH-like. Often, only the lowest HH subband is occupied, i.e.,
the occupied subband states are predominantly made of $m=\pm 3/2$
states assuming a spin quantization axis perpendicular to the 2D
plane. It is this fact that lies at the heart of several anomalous
properties of 2D hole systems that we are going to discuss in
subsequent sections. Of course, at magnetic field $B=0$ the $m=+3/2$
and $m=-3/2$ states contribute with equal weight, i.e., as expected
for a paramagnetic material, the effects discussed here do not give
rise to a permanent magnetic moment.

For a more explicit and detailed discussion of spin orientation we
need to use the Hamiltonian appropriate for these hole systems.
Generally, the $j=3/2$ hole states in the topmost valence band of
cubic semiconductors like GaAs are described by the Luttinger
Hamiltonian~\cite{lut56}
\begin{equation}
  \label{eq:lutt_lutt}
  \begin{array}[b]{r@{}l}
 \mathcal{H}_L = \Ds -\frac{\hbar^2}{2 m_0}
 \bigl\{ \gamma_1 k^2 &{}
 - 2 \gamma_2 \left[\left(J_x^2 - \frack{1}{3}J^2\right) k_x^2
   + \mathrm{cp} \right] \\[1ex] & {}
 - 4 \gamma_3 \left[ \left\{ J_x, J_y \right\} k_x k_y +
   \mathrm{cp}\right] \bigr\},
\end{array}
\end{equation}
where $m_0$ is the mass of free electrons, $\gamma_1$, $\gamma_2$,
and $\gamma_3$ are the Luttinger parameters, $J_i$ are the $4\times
4$ matrices for angular momentum $j=3/2$ (see, e.g., Ref.\
\cite{win03}), we have $\{ A, B \} = \frac{1}{2} (AB + BA)$ and cp
denotes cyclic permutation. In Eq.\ (\ref{eq:lutt_lutt}) we
neglected the terms linear and cubic in $k$ which contribute to the
spin splitting due to bulk inversion asymmetry \cite{win03}. We
remark that for the effects discussed here these terms are of minor
importance. Using an explicit matrix notation, $\mathcal{H}_L$ can
be written in the form~\cite{bal70}
\begin{equation}
  \label{eq:lutt_bl}
  \mathcal{H}_L = \left( \begin{array}{cccc}
      P+Q & L & M & 0 \\
      L^\ast & P-Q & 0 & M \\
      M^\ast & 0 & P-Q & -L \\
      0 & M^\ast & -L^\ast & P+Q
    \end{array} \right),
\end{equation}
where
\begin{subequations}
  \begin{eqnarray}
    P & = & - \frac{\hbar^2}{2m_0} \, \gamma_1 \, k^2 \\
    Q & = & \frac{\hbar^2}{2m_0} \, \gamma_2 \, (2k_z^2 - k_x^2 - k_y^2) \\
    L & = & \frac{\hbar^2}{2m_0} \, 2\sqrt{3} \gamma_3 \, (k_x - ik_y) k_z \\
    M & = & \frac{\hbar^2}{2m_0} \, \sqrt{3}
    \left[ \gamma_2 (k_x^2-k_y^2)
      - 2i \gamma_3 \, k_x k_y \right].
  \end{eqnarray}
\end{subequations}
Here we have expressed $J_i$ and $\mathcal{H}_L$ in a basis of
$j=3/2$ angular momentum eigenfunctions in the order $m=+3/2$, $+1/2$,
$-1/2$, and $-3/2$.

The notation used in Eq.\ (\ref{eq:lutt_lutt}) reflects the cubic
symmetry of the crystal structure. An alternative formulation of the
Luttinger Hamiltonian was proposed by Lipari and Baldereschi
\cite{lip70}
\begin{equation}
  \label{eq:lutt_lb}
  \mathcal{H}_L = - \frac{\hbar^2}{2m_0} \gamma_1 k^2
  - \frac{\hbar^2}{2m_0} \bar{\gamma} 
  \left[ {\textstyle\frac{5}{2}} k^2 - 2
    (\kk \cdot \vek{J})^2 \right]
  + \mathcal{H}_\mathrm{c},
\end{equation}
where $\bar{\gamma} \equiv (2\gamma_2 + 3\gamma_3) / 5$ and $\vek{J}
= (J_x,J_y,J_z)$. The first two terms in Eq.\ (\ref{eq:lutt_lb})
have spherical symmetry with the first term being diagonal in spin
while the second term can be interpreted as a spherical spin-orbit
coupling within the $j=3/2$ space, see also Eq.\
(\ref{eq:lutt_spher}) below. Finally, $\mathcal{H}_\mathrm{c}$
represents the anisotropic terms with cubic symmetry \cite{lip70}
which will be given in Eq.\ (\ref{eq:luttcub}) below. Usually the
terms in $\mathcal{H}_\mathrm{c}$ are small. Neglecting these terms
corresponds to the \emph{spherical approximation} $\mathcal{H}_s$ of
$\mathcal{H}_L$. Using for $\mathcal{H}_s$ an explicit matrix
notation as in Eq.\ (\ref{eq:lutt_bl}) we get $P_s = P$ and
\begin{subequations}
  \begin{eqnarray}
    Q_s & = & \frac{\hbar^2}{2m_0} \, \bar{\gamma} \,
     (2k_z^2 - k_x^2 - k_y^2) \\
    L_s & = & \frac{\hbar^2}{2m_0} \, 2\sqrt{3} \bar{\gamma} \, k_- k_z \\
    M_s & = & \frac{\hbar^2}{2m_0} \, \sqrt{3} \bar{\gamma} \, k_-^2,
  \end{eqnarray}
\end{subequations}
where $k_\pm \equiv k_x \pm i k_y$. In the spherical approximation,
the energy dispersions for the HH and LH states are
\begin{equation}
  \label{eq:spher_disp}
  E_\mathrm{LH/HH} (\kk) = - \frac{\hbar^2}{2m_0} \,
  (\gamma_1 \pm 2\bar{\gamma}) \: k^2.
\end{equation}

For quasi-2D systems it is often advantageous to use an alternative
decomposition $\mathcal{H}_L = \mathcal{H}_\mathrm{ax} +
\mathcal{H}_\mathrm{c}'$, where $\mathcal{H}_\mathrm{ax}$ has axial
symmetry with the symmetry axis $\vek{n}$ chosen perpendicular to
the 2D plane. Neglecting $\mathcal{H}_\mathrm{c}'$ corresponds to
the \emph{axial approximation} \cite{tre79, win03}. For $\vek{n}$
parallel to $[001]$ we obtain for $\mathcal{H}_\mathrm{ax}$, using an
explicit matrix notation as in Eq.\ (\ref{eq:lutt_bl}),
\begin{subequations}
  \begin{eqnarray}
    P_\mathrm{ax} & = & P, \qquad
    Q_\mathrm{ax} = Q, \qquad L_\mathrm{ax} = L \\
    M_\mathrm{ax} & = & \frac{\hbar^2}{2m_0} \, \sqrt{3} \,
    \frac{\gamma_2+\gamma_3}{2} \, k_-^2.
  \end{eqnarray}
\end{subequations}
Explicit expressions for other crystallographic orientations of
$\vek{n}$ are given in Refs.\ \cite{tre79, win03}. For quasi-2D
systems the advantage of the axial approximation over the spherical
approxiamtion lies in the fact that $\mathcal{H}_\mathrm{ax}$
captures the most important physics of different crystallographic
directions $\vek{n}$ while both $\mathcal{H}_\mathrm{s}$ and
$\mathcal{H}_\mathrm{ax}$ yield a rotational symmetry with respect
to the axis $\vek{n}$.

We can readily see from Eq.\ (\ref{eq:lutt_bl}) that $\mathcal{H}_L$
becomes diagonal for $k_x=k_y=0$, i.e., subband states at $k_\|=0$
are either pure HH ($m=\pm 3/2$) or pure LH ($m=\pm 1/2$) states. As
discussed above, in a coordinate frame where the $z$ axis points in
a crystallographic direction other than the high-symmetry directions
[001] or [111], $\mathcal{H}_L$ contains off-diagonal terms
proportional to $k_z^2$ so that even at the subband edge $k_\|=0$ we
get a mixing of HH and LH states.

\section{Anisotropic Zeeman Splitting}
\label{sec:zeeman}

For spin-1/2 electron systems it is well-known that the Zeeman
splitting, i.e., the response of the electron's spin to a magnetic
field $\vek{B}$, is usually essentially independent of the
orientation of $\vek{B}$ \cite{fan68}. For electrons, the
differences between the in-plane and perpendicular effective Land\'e
factors $g^\ast_\|$ and $g^\ast_\perp$ are most pronounced in narrow
GaAs-AlGaAs quantum wells, where $g^\ast_\|$ and $g^\ast_\perp$
change sign and thus cross zero as a function of well width, as
discussed in Refs.\ \cite{ivc92, kal92}. The in-plane anisotropy of
the in-plane $g^\ast_\|$ in low-symmetry geometries was discussed in
Ref.\ \cite{kal93}, see also Ref.\ \cite{win03}.

The situation in HH systems is qualitatively different. As suggested
by Fig.~\ref{fig:hspin}, the response of a HH system to an in-plane
magnetic field $B_\|$ is suppressed because the effect of $B_\|$
competes with the rather rigid spin orientation induced by the HH-LH
splitting. A perpendicular magnetic field $B_\perp$, on the other
hand, is compatible with the $B=0$ spin orientation of the $m=\pm
3/2$ HH states so that we can have a large Zeeman splitting. The
resulting anisotropy of $g^\ast$ was first discussed in
Ref.~\cite{kes90}.

A more quantitative discussion needs to be based on the Luttinger
Hamiltonian $\mathcal{H}_L$ with the Zeeman term $\mathcal{H}_Z =
-2\bohrmag \,\kappa \, \vek{B}\cdot\vek{J}$ added to it. Here
$\kappa$ is the isotropic $g$ factor. We neglect the anisotropic
Zeeman term because its prefactor (often denoted $q$ \cite{lut56})
is usually much smaller than $\kappa$. Using an explicit matrix
notation, the Zeeman term reads
\begin{equation}
  \label{eq:lutt_zeeman}
  \mathcal{H}_Z =
  \arraycolsep 0.1em
  \newcommand{\wddzz}{\frac{\sqrt{3}}{2}}
  \renewcommand{\arraystretch}{1.4}
 -2 \kappa \bohrmag \left(\!\begin{array}{cccc}
  \frac{3}{2}B_z & \wddzz B_- & 0 & 0 \\
  \wddzz B_+ & \frac{1}{2}B_z & B_- & 0 \\
  0 & B_+ & -\frac{1}{2}B_z & \wddzz B_- \\
  0 & 0 & \wddzz B_+ & -\frac{3}{2}B_z 
  \end{array}\!\right) ,
\end{equation}
where $B_\pm \equiv B_x \pm iB_y$. We can immediately read off from
Eq.\ (\ref{eq:lutt_zeeman}) that a perpendicular magnetic field
$B_z$ gives rise to a Zeeman splitting $\Delta E_\mathrm{HH} = 6
\bohrmag \kappa B_z$ of the HH states whereas for LH states we get
$\Delta E_\mathrm{HH} = 2 \bohrmag \kappa B_z$. Also, we see from
Eq.\ (\ref{eq:lutt_zeeman}) that, in the presence of an in-plane
magnetic field $\vek{B}_\| = (B_x,B_y,0)$, the Zeeman term
$\mathcal{H}_Z$ couples the two LH states with $\Delta E_\mathrm{LH}
= 4 \bohrmag \kappa B_\|$, and it couples the HH states to the LH
states~\cite{gol93}. But there is no direct coupling between the HH
states proportional to $\kappa$, so that the Zeeman splitting of HH
states in an in-plane magnetic field is suppressed \cite{kes90}.

A quantitative analysis based on perturbation theory shows
\cite{win03} that, for $k_\|=0$ and neglecting the cubic terms
$\mathcal{H}_\mathrm{c}$, there is no Zeeman splitting of HH states
linear in $B_\|$, but the lowest-order splitting is proportional
to~$B_\|^3$
\begin{subequations}
  \label{eq:zeeman_par}
  \begin{equation}
    \label{eq:zeeman_kub}
    \Delta E_{Z\|} \propto \frac{B_\|^3}{\Delta_{11}^{hl}},
  \end{equation}
  where $\Delta_{\lambda\lambda'}^{\nu\nu'} \equiv E_\lambda^\nu -
  E_{\lambda'}^{\nu'}$ with $E_\lambda^h$ and $E_\lambda^l$ the
  energy of the $\lambda$th HH and LH subband, respectively.
  Similarly, we obtain for $k_\|>0$
  \begin{equation}
    \label{eq:zeeman_kpar}
    \Delta E_{Z\|} \propto \frac{k_\|^2 \, B_\|}{\Delta_{11}^{hl}}.
  \end{equation}
\end{subequations}
As expected from Fig.~\ref{fig:hspin}, the energy denominators in
Eq.\ (\ref{eq:zeeman_par}) reflect the competition between the spin
orientation perpendicular to the 2D plane induced by the subband
confinement and the (generally weaker) effect of the in-plane
magnetic field $B_\|$ that tends to orient the spins in-plane. The
$k_\|$ dependence in Eq.\ (\ref{eq:zeeman_kpar}) originates in the
HH-LH coupling at $k_\|>0$. It follows from Eq.\
(\ref{eq:zeeman_par}) that we can tune the Zeeman splitting of HH
systems in an in-plane magnetic field if we change the confinement
in $z$ direction by means of front and/or back gates. Alternatively,
one can also change the HH-LH splitting by means of strain.

\begin{figure}[tbp]
\centerline{\includegraphics[width=60mm]{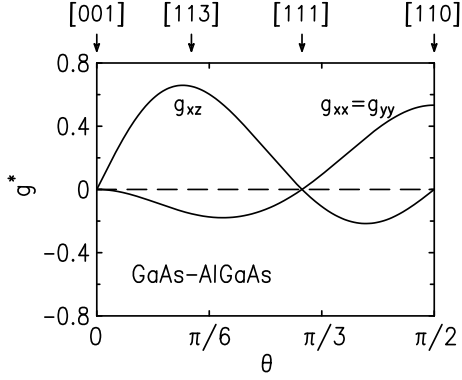}}
\caption[]{\label{fig:gfak_anis_gaas} Anisotropic effective $g$
factor $g^\ast$ of the $h_1$ subband for a GaAs--AlGaAs QW as a
function of $\theta$, the angle between [001] and the growth
direction. Here $\hat{\vek{x}} \parallel [nn\overline{(2m)}]$,
$\hat{\vek{y}} \parallel [\bar{1}10]$ and
$\hat{\vek{z}} \parallel [mmn]$.}
\end{figure}

When taking into account the cubic terms $\mathcal{H}_\mathrm{c}$ in
the Luttinger Hamiltonian, the Zeeman splitting linear in $B_\|$
($k_\|=0$) still vanishes for the high-symmetry surfaces (001) and
(111). For other surfaces, we get a $B_\|$-linear splitting the
magnitude of which depends on the in-plane orientation of
$\vek{B}_\|$ relative to the crystal axes \cite{win00b}. For an
infinitely deep rectangular QW grown on an $(mmn)$ surface we obtain
in second order perturbation theory the following Zeeman terms
acting within the space of the topmost HH subband
\begin{subequations}
  \label{eq:zeeman_hh}
  \begin{equation}
    \mathcal{H}^\mathrm{HH}_{[mmn]} = 
    \frac{\bohrmag}{2} \left( g_{xx}^\ast B_x \sigma_x
      + g_{xz}^\ast B_x \sigma_z +  g_{yy}^\ast B_y \sigma_y \right)
  \end{equation}
  where
  \begin{eqnarray}
    g_{xx}^\ast & = & g_{yy}^\ast = - 6 \mathcal{K}
    \left(2-3\sin^2\theta\right) \sin^2\theta \\[1.5ex]
    g_{xz}^\ast & = & 12 \mathcal{K}
    \left(2-3\sin^2 \theta\right) \sin\theta \cos\theta
  \end{eqnarray}
\end{subequations}
Here the $x$ direction corresponds to $[nn\overline{(2m)}]$, $y$
corresponds to $[\overline{1}10]$ and $\theta$ is the angle between
$[mmn]$ and $[001]$, i.e.\ $\theta = \arccos (n / \sqrt{2m^2+n^2})$.
We obtain for the coefficient $\mathcal{K}$
\begin{subequations}
  \label{eq:gfak}
\begin{equation}
\mathcal{K} = 
\frac{\kappa (\gamma_3 - \gamma_2)}
{4 \left[\left(1-\zeta\right) \gamma_2 + \zeta\gamma_3 \right]}
\end{equation}
where
\begin{equation}
\zeta = \sin^2 \theta  \left[3 - \frack{9}{4} \sin^2 \theta  \right] .
\end{equation}
\end{subequations}
Note that these expressions are independent of the width of the QW.
The importance of Eq.\ (\ref{eq:zeeman_hh}) lies in the fact that
the topmost subband in an (unstrained) QW is an HH subband so that
often only this subband is occupied. In
Fig.~\ref{fig:gfak_anis_gaas} we show $g_{xx}^\ast = g_{yy}^\ast$
and $g_{xz}^\ast$ for the topmost HH subband in a GaAs--AlGaAs QW as
a function of the angle $\theta$. Figure~\ref{fig:gfak_anis_gaas}
demonstrates that $g^\ast$ can be very anisotropic. For example, for
the growth direction [113], $g^\ast_{xz}$ is more than a factor of
four larger than $g_{xx}^\ast = g_{yy}^\ast$. For comparison, we
remark that for the GaAs system considered in
Fig.~\ref{fig:gfak_anis_gaas} we have $g_{zz}^\ast = 6 \kappa
\approx 7.2$.
Equations (\ref{eq:zeeman_hh}) and (\ref{eq:gfak}) are applicable to
a wide range of cubic semiconductors with results qualitatively very
similar to Fig.~\ref{fig:gfak_anis_gaas}.

\begin{figure}[t]
\centerline{\includegraphics[width=85mm]{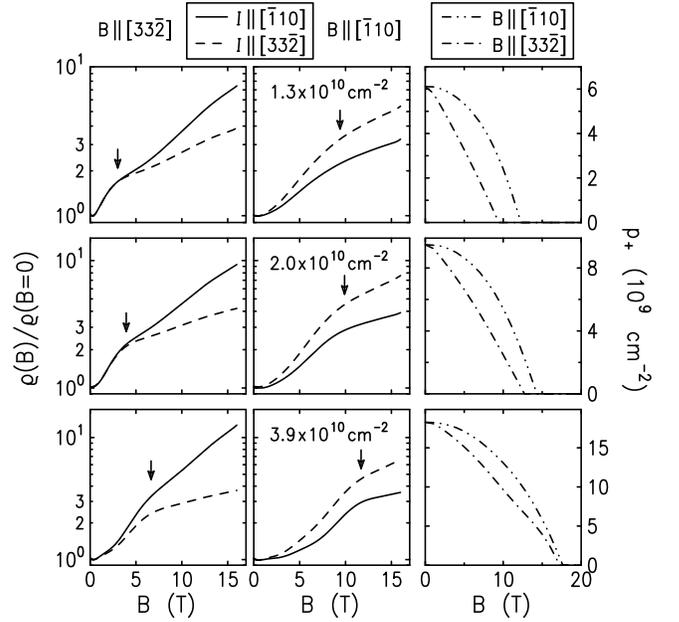}}
\caption[]{\label{pic:gfak_exp}Left and central panels: Fractional
change in resistivity $\rho(B_\|)/\rho(0)$ due to an in-plane $B$,
measured at $T=0.3$~K in a GaAs 2D hole system grown on a (113)
substrate, for different directions of $\vek{B}_\|$ and $\vek{I}$
and different 2D densities as indicated. The arrows mark $B^\ast$ as
defined in the text. Right panel: Calculated density $p_+$ in the
upper spin subband as a function of $B_\|$.}
\end{figure}

The anisotropic Zeeman splitting (\ref{eq:zeeman_hh}) has been
studied experimentally by measuring the magnetoresistance of a
high-mobility 2D hole system as a function of in-plane magnetic
field $B_\|$ \cite{win00b}. The sample was a 200 {\AA} wide
Si-modulation doped GaAs QW grown on (113)A GaAs substrate. The left
two panels of Fig.~\ref{pic:gfak_exp} show the resistivity~$\rho$
measured as a function of $B_\|$ for different directions of
$\vek{B}_\|$ and current~$\vek{I}$ and for three different
densities. For easier comparison we have plotted the fractional
change $\rho (B_\|) / \rho(0)$. It can be seen that $\log (\rho)$
shows a change in slope at a value of $B_\|$ we call $B^\ast$. In
Fig.~\ref{pic:gfak_exp} $B^\ast$ is marked by arrows. This
magnetoresistance feature is related to a spin-subband depopulation
and the resulting changes in subband mobility and intersubband
scattering as $B_\|$ is increased \cite{oka99, pap00a}. It is
remarkable that $B^\ast$ for the $\vek{B} \parallel [33\bar{2}]$
traces is several Tesla smaller than for the $[\bar{1}10]$ traces,
but it is independent of the direction of $\vek{I}$. This is strong
evidence for the anisotropy of the in-plane $g^\ast$. The
experimentally observed anisotropy is qualitatively consistent with
our self-consistently calculated results for the density $p_+$ of
the upper spin subband as a function of $B_\|$, shown in the right
panel of Fig.~\ref{pic:gfak_exp}. The density $p_+$ decreases much
faster for $\vek{B} \parallel [33\bar{2}]$ than for
$\vek{B} \parallel [\bar{1}10]$, in agreement with
Fig.~\ref{fig:gfak_anis_gaas}. The findings have been confirmed by
Shubnikov-de Haas measurements probing directly the depopulation of
the minority spin subband for different orientations of $\vek{B}_\|$
\cite{tut01}.

In Fig.~\ref{pic:gfak_exp} the measured $B^\ast$ is significantly
smaller than the calculated $B_\|$ for a complete depopulation of
the upper spin-subband. We note that for our low-density samples it
can be expected that $g^\ast$ is enhanced due to the exchange
interaction and the spin polarization caused by $B_\|$ \cite{fan68,
kwo94, oka99}. These many-particle effects were not taken into
account in our self-consistent calculations. However, they do not
qualitatively affect the anisotropy of $g^\ast$ \cite{win00b}. See
Ref.~\cite{win05} for a more detailed discussion of
exchange-correlation effects in low-density hole systems on a (113)
surface.

\section{Rashba Spin Splitting in 2D Hole Systems}
\label{sec:rashba}

If in a solid the spatial inversion symmetry is broken, we can have
a spin splitting of the electron and hole states even in the absence
of a magnetic field $B$. In quasi-2D semiconductor structures, the
bulk inversion asymmetry (BIA) of the underlying crystal structure
(e.g., a zinc blende structure), and the structure inversion
asymmetry (SIA) of the confining potential are usually the dominant
contributions to the $B=0$ spin splitting \cite{ros89}. While BIA
is fixed, the so-called Rashba spin splitting \cite{byc84} due to
SIA can be tuned by means of external gates that change the electric
field $\Ee$ in the sample \cite{nit97, pap99}.

Here we want to focus on Rashba spin splitting. In quasi-2D electron
systems it is described by the term
\begin{equation}
\label{eq:rashba}
\mathcal{H}_R^\mathrm{e} = \alpha \Ee_z 
        i \left( k_- \sigma_+ - k_+ \sigma_-\right)
   =  \alpha \Ee_z \left(
  \begin{array}{cc} 0 & i k_- \\ -i k_+ & 0 \end{array}
  \right) \,,
\end{equation}  
where $\sigma_\pm \equiv \frac{1}{2}(\sigma_x \pm i\sigma_y)$,
$\alpha$ is a material-dependent prefactor and $\Ee_z$ is an
(effective) electric field that characterizes the broken inversion
symmetry of the sample. Often it is instructive to write the Rashba
term (\ref{eq:rashba}) in the form of a Zeeman term \cite{win04}
\begin{equation}
  \label{eq:rash_effB}
  \mathcal{H}_R^\mathrm{e} = {\textstyle\frac{1}{2}} \:
  \vekc{B} (\kk_\|) \cdot \vek{\sigma}
\end{equation}
where $\vekc{B} (\kk_\|) = 2\alpha\Ee_z (k_y, -k_x, 0)$ is an
effective magnetic field. It follows from the defintion of $\vekc{B}
(\kk_\|)$ that $\vekc{B}$ orients the electron spin in the plane of
the 2D structure, perpendicular to the wave vector $\kk_\|$.

For $j=3/2$ hole systems we need to replace the $2\times 2$ Pauli
matrices $\sigma_x$ and $\sigma_y$ in Eq.\ (\ref{eq:rashba}) or
(\ref{eq:rash_effB}) by the corresponding $4\times 4$ matrices $J_x$
and $J_y$ for angular momentum $j=3/2$, while the definition of the
effective magnetic field $\vekc{B} (\kk_\|)$ is valid also for hole
systems, i.e., exactly as in electron systems, the field $\vekc{B}
(\kk_\|)$ tends to orient the hole spins in-plane. Here, many
arguments on Zeeman splitting of hole states in an in-plane magnetic
field carry over to the splitting in the effective field
$\vekc{B}(\kk_\|)$. In particular, similar to Eq.\
(\ref{eq:zeeman_par}), Rashba splitting of hole systems is
proportional to $k_\|^3$ which is qualitatively different from the
$k_\|$-linear Rashba spin splitting of electron systems. Within the
two-dimensional subspace of one HH subband, the Rashba term
describing the cubic splitting is of the form
\begin{equation}
\label{eq:rashba_HH}
\mathcal{H}_R^\mathrm{HH} = \beta^h \Ee_z 
        i \left( k_+^3 \sigma_- - k_-^3 \sigma_+\right) \,.
\end{equation}
In electron systems, the coefficient $\alpha$ depends only
on the material, but it is essentially independent of the geometry
of the quasi-2D system \cite{win03}. In HH systems, on the other
hand, we get in third-order perturbation theory for the first HH
subband
\begin{equation}
  \label{eq:rash_HH_fak}
  \beta^h_1 = \frac{e\hbar^4}{m_0^2} \,a\, \gamma_3 (\gamma_2+\gamma_3)
  \bigg[ \frac{1}{\Delta_{11}^{hl}} 
  \bigg(\frac{1}{\Delta_{12}^{hl}} 
     - \frac{1}{\Delta_{12}^{hh}} \bigg)
     + \frac{1}{\Delta_{12}^{hl} \, \Delta_{12}^{hh}} \bigg]
\end{equation}
where $a=64/(9\pi^2)$ for an infinitely deep, rectangular QW, i.e.,
as expected from Fig.~\ref{fig:hspin}, the prefactor $\beta^h_1$
depends on the HH-LH splitting which, in turn, depends on the
geometry of the quasi-2D system. The detailed comparison of Eqs.\
(\ref{eq:zeeman_par}) and (\ref{eq:rash_HH_fak}) reveals a subtle
difference between Zeeman splitting and Rashba spin splitting in HH
systems. While Eq.\ (\ref{eq:zeeman_par}) can be derived in
second-order perturbation theory, each term in Eq.\
(\ref{eq:rash_HH_fak}) contains two energy denominators $1/\Delta$.
This is due to the fact that the effective field $\vekc{B}$ depends
on the wave vector $\kk_\|$ whereas $\vek{B}$ is independent of
$\kk_\|$.

The functional form of the HH Rashba coefficient $\beta^h$ gives
rise to a remarkable difference between Rashba spin splitting in
electron and hole systems. It is well-known that Rashba spin
splitting in electron systems is roughly linearly proportional to
the electric field $\Ee_z$ that characterizes the inversion
asymmetry of the confining potential \cite{ros89, nit97, pap99,
win00a, las85, and97, win03}. This reflects the fact that the
coefficient $\alpha$ in Eq.\ (\ref{eq:rashba}) is essentially
independent of the geometry of the quasi-2D electron systems. In
hole systems the $\Ee_z$ dependence of spin splitting can be
reversed, i.e., a large field $\Ee_z$ can give rise to a small spin
splitting and vice versa \cite{hab04a}. This is due to the fact that
the HH-LH splittings $\Delta^{hl}$ in Eq.\ (\ref{eq:rash_HH_fak})
depend on the geometry of the quasi-2D system that can be tuned by
means of $\Ee_z$. Indeed, the implicit dependence of $\beta^h_1$ on
$\Ee_z$ can be such that not only, it cancels the explicit $\Ee_z$
dependence in Eq.\ (\ref{eq:rashba_HH}), but it can even result in
an inverse dependence of spin splitting on $\Ee_z$.

\begin{figure}[tbp]
\centerline{\includegraphics[width=68mm]{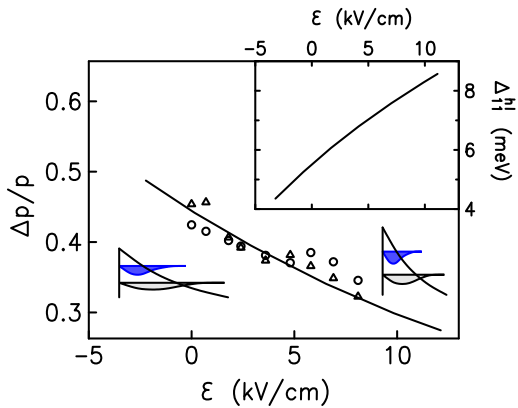}}
\caption[]{\label{fig:arash}Spin splitting $\Delta p/p$ for a
quasi-2D hole system in a (001) GaAs-Al$_{0.3}$Ga$_{0.7}$As single
heterojunction at constant total density $p = 1.84 \times
10^{11}$~cm$^{-2}$, where the asymmetry was tuned by applying an
electric field $\Ee_\perp$ perpendicular to the quasi-2D system
\cite{hab04a}. The inset shows the calculated HH-LH splitting
$\Delta^{hl}_{11}$ that decreases as a function of $\Ee_z$.}
\end{figure}

Experimentally, the spin splitting gives rise to a difference
$\Delta p = p_+ - p_-$ between the densities $p_+$ and $p_-$ in the
two spin subbands \cite{win03}. As an example, we show in
Fig.~\ref{fig:arash} the measured and calculated spin splitting
$\Delta p/p$ for a quasi-2D hole system in a Be-doped (001)
GaAs-Al$_{0.3}$Ga$_{0.7}$As single heterojunction at constant total
density $p = p_+ + p_- = 1.84 \times 10^{11}$~cm$^{-2}$, where the
asymmetry was tuned by applying an electric field $\Ee_z$
perpendicular to the quasi-2D system using front and back gates
\cite{hab04a}. It can be seen that the spin splitting is reduced
when $\Ee_z$ is increased which is opposite to the behavior observed
in electron systems \cite{nit97}.

\section{Multipole Expansion of the Spin Density Matrix}
\label{sec:muexp}

We can gain a more systematic understanding of the unusual
spin-dependent phenomena in hole systems from the spin density
matrix which is the fundamental object providing a complete
description of the system. Neglecting the orbital degrees of
freedom, spin-1/2 electron systems are characterized by a $2\times
2$ spin density matrix $\rho$, whereas in spin-3/2 hole systems
$\rho$ becomes a $4\times 4$ matrix. The dominant character of the
occupied eigenstates of hole systems depends on the quantization
axis of the underlying $j=3/2$ basis functions. Obviously,
observable quantities such as the spin polarization may not depend
on this choice. Therefore, it is necessary to formulate the spin
density matrix in a way such that observable quantities can be
calculated independent of the particular choice for the basis
functions that are used. Using the theory of invariants
\cite{bir74,win03} it is possible to derive an invariant
decomposition of the spin density matrix of $j=3/2$ hole systems
that can be interpreted as a multipole expansion \cite{win04b}.
These multipoles indeed have the desired property that they can be
evaluated and interpreted independent of the particular choice of
basis functions.

Quite generally, neglecting small terms with cubic symmetry, the
spin density matrix ${\rho}$ for systems with spin $j$ can be
decomposed as follows \cite{win04b}
\begin{subequations}
  \label{eq:rhodecomp}
  \begin{eqnarray}
    {\rho} & = & \sum_{\fj = 0}^{2j} {\vek\rho}_{\fj}\cdot\vek{M}_{\fj}\\
    & \equiv & \rho_0 \, M_0  + \vekc{S}\cdot\vek{M}_{1}
    + \vekc{Q}\cdot\vek{M}_{2} + \vekc{O}\cdot\vek{M}_{3}
    + \ldots . \hspace*{2em}
  \end{eqnarray}
\end{subequations}
The $2\fj + 1$ quantities $\vek{M}_{\fj}$, which we refer to as
multipoles, are spherical tensors that transform according to the
irreducible representations $\mathcal{D}_\fj$ of the point group
SU(2). They are analogs of the vector of Pauli spin matrices
familiar in electron systems. Each $\vek{M}_{\fj}$ is a $(2\fj +
1)$-dimensional vector with components $M_{\fj \fm}$, $- \fj \le
{\fm} \le \fj$, which are $(2j + 1)\times(2j + 1)$ matrices. These
matrices are orthonormal in the sense that $\trace (M^\dag_{\fj \fm}
\, M_{\fj'\fm'}) = \delta_{\fj\fj'}\delta_{\fm\fm'}$. They can be
calculated using standard angular momentum theory \cite{edm60}.
$M_{\fj \fm}$ for spin-$1/2$ and spin-$3/2$ systems have been
tabulated in Ref.~\cite{win04b}. The multipole moments
${\vek\rho}_{\fj}$ give the weight of each multipole in the density
matrix. Each ${\vek\rho}_{\fj}$ is a $(2\fj + 1)$-dimensional vector
with scalar components $\rho_{\fj \fm}$. The components $\rho_{\fj
\fm}$ of the moments $\vek{\rho}_\fj$ can be obtained from
\begin{equation}
\rho^\ast_{\fj \fm} = \trace (M_{\fj \fm} \, \rho),
\end{equation}
and the magnitude of
each moment is given by
\begin{equation}
|\rho_\fj|^2 = \sum_{\fm} (-1)^{\fm} \, \rho_{\fj \fm} \, \rho_{\fj,-\fm}.
\end{equation}
In Eq.\ (\ref{eq:rhodecomp}) one can thus interpret the vectors
$\vek{M}_\fj$ as multipole operators that ``measure'' the moments
$\vek{\rho}_\fj$ of the system. For simplicity we assume that all
orbital degrees of freedom are integrated over, so that ${\rho}$
depends only on the spin indices and does not depend on the wave
vector $\kk$. Nevertheless, the coupling of spin and $\kk$ is
present in the Hamiltonian through the spin-orbit interaction. We
note that the dot product for spherical tensors appearing in Eq.\
(\ref{eq:rhodecomp}) is defined in Ref.\ \cite{edm60}.

The moments ${\vek\rho}_{\fj}$ provide a set of \emph{independent}
parameters characterizing the matrix ${\rho}$ \cite{win04b}. The
monopole $\rho_0$ is identified with the carrier density, while the
dipole $\vek{\rho}_1 \equiv \vekc{S}$ corresponds to the spin
polarization or Bloch vector at $B>0$. These are the only moments
present in the density matrix of conduction electrons. In $j=3/2$
hole systems, a quadrupole $\vek{\rho}_2 \equiv \vekc{Q}$ is also
present, which reflects the splitting between the HH and LH states
discussed in the preceding sections. The octupole $\vek{\rho}_3
\equiv \vekc{O}$ is a unique feature of $j=3/2$ systems at $B>0$.

As discussed in Sec.~\ref{sec:luttham}, the dynamics of $j=3/2$ hole
systems is characterized by the $4 \times 4$ Luttinger Hamiltonian
$\mathcal{H}_L$. Obviously, the spherical tensor operators discussed
here provide the most natural language to formulate the spherical
approximation $\mathcal{H}_s$ of $\mathcal{H}_L$, i.e., instead of
Eq.\ (\ref{eq:lutt_lb}) we may write $\mathcal{H}_s$ analogous to
Eq.\ (\ref{eq:rhodecomp}) in the form \cite{lip70, suz74}
\begin{subequations}
  \label{eq:lutt_spher}
  \begin{eqnarray}
    \mathcal{H}_s & = & \sum_{\fj=0}^3 \mathcal{H}_{\fj}
  = \sum_{\fj=0}^3 a_{\fj} \, \vekc{K}_{\fj} \cdot \vekc{M}_{\fj}
  \label{eq:lutt_spherg} \\
     & = & - \frac{\hbar^2 \, \gamma_1}{m_0} \, \vekc{K}_0 \cdot \vek{M}_0 
    - 2\sqrt{5} \, \kappa \bohrmag \, \vekc{K}_1 \cdot \vek{M}_1
    \nonumber \\[2ex] & & \label{eq:lutt_sphere}
    + \sqrt{6}\, \frac{\hbar^2 \, \bar{\gamma}}{m_0}
    \, \vekc{K}_2 \cdot \vek{M}_2
    + \zeta \, \vekc{K}_3 \cdot \vek{M}_3
    + \mathcal{H}_\mathrm{c} \:. \hspace*{2em}
  \end{eqnarray}
\end{subequations}
The tensor operators $\vekc{K}_\fj$ (like the vectors $\vek{M}_\fj$) are
tabulated in Ref.~\cite{win04b}. The first term in Eq.\
(\ref{eq:lutt_sphere}), which we interpret as a monopole, is equal
to the first term in Eq.\ (\ref{eq:lutt_lb}). The second term in
Eq.\ (\ref{eq:lutt_sphere}), which corresponds to a dipole, is the
(spherically symmetric) Zeeman term (\ref{eq:lutt_zeeman}). The
third term in Eq.\ (\ref{eq:lutt_sphere}), which corresponds to a
quadrupole, is equal to the second term in (\ref{eq:lutt_lb}). This
term is responsible for the HH-LH splitting. The fourth term in Eq.\
(\ref{eq:lutt_sphere}) corresponds to an octupole, which depends on
$\kk$ and $\vek{B}$. The prefactor $\zeta$ of this term is typically
very small so that normally it can be neglected~\cite{suz74,
win04b}. Obviously $\mathcal{H}_1 = \mathcal{H}_3 = 0$ for zero
magnetic field. We note that in Eq.\ (\ref{eq:lutt_lb}) we
considered only the terms with the lowest order of $\kk$ and
$\vek{B}$ so that we have $\mathcal{K}_0 = k^2$ and $\vekc{K}_1 =
\vek{B}$. Higher-order terms as well as, e.g., strain-induced terms
can be classified in the same way. For example, the most important
strain-induced term is a quadrupole field (independent of $\kk$
\cite{win04b}). Finally, the cubic term $\mathcal{H}_\mathrm{c}$
reads
\begin{equation}
  \label{eq:luttcub}
  \begin{array}[b]{@{}r@{}l@{}}
  \mathcal{H}_\mathrm{c} = \Ds - \frac{\hbar^2}{2m_0} & \, \sqrt{6} \,
  (\gamma_3 - \gamma_2) \, \Bigl\{
    \left[\vekc{K}_2\times\vek{M}_2\right]_{4,-4} \\[2ex] & 
    + {\textstyle\frac{\sqrt{70}}{5}}
    \left[\vekc{K}_2\times\vek{M}_2\right]_{4,0}
    + \left[\vekc{K}_2\times\vek{M}_2\right]_{4,4}
  \Bigr\} ,
\end{array}
\end{equation}
where the product of spherical tensors is defined in
Ref.~\cite{edm60}. Our prefactor of $\mathcal{H}_\mathrm{c}$ differs
slightly from the one given in Ref.\ \cite{lip70} due to the
different normalization of the tensor operators $\vekc{K}_2$ and
$\vek{M}_2$ adopted from Ref.\ \cite{win04b}.

\section{Multipole Moments Induced by a Magnetic Field}
\label{sec:muma}

\begin{figure}[tbp]
\centerline{\includegraphics[width=68mm]{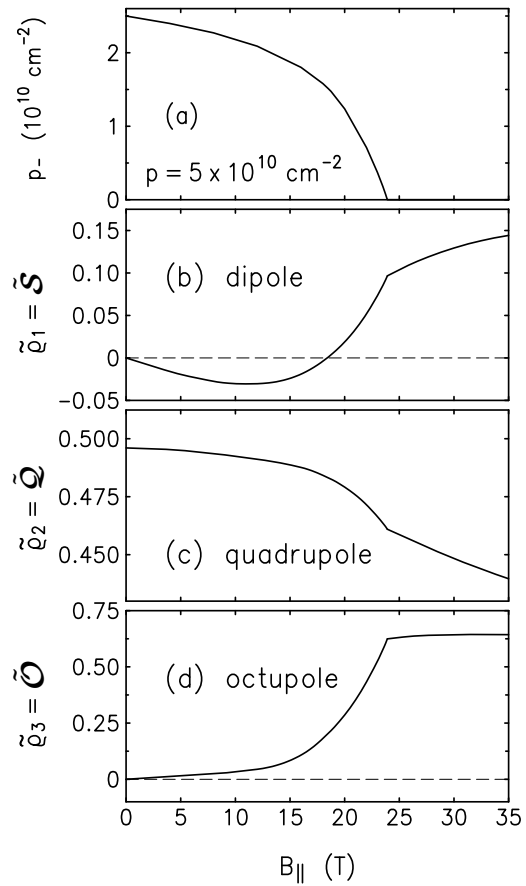}}
\caption{\label{fig:mumo}(a) Spin subband density $p_-$ of the HH
minority spin subband and (b-d) normalized multipole moments
$\tilde{\rho}_i$ as a function of the in-plane magnetic field $B_\|$
calculated self-consistently for a symmetric (100)
GaAs-Al$_{0.3}$Ga$_{0.7}$As quantum well with hole density $p = 5
\times 10^{10}$~cm$^{-2}$ and well width $w=150$~{\AA}. Adapted from
Ref.\ \cite{win04b}.}
\end{figure}

To gain a qualitative understanding of the relevance of the spin
multipoles $\vek{\rho}_\fj$, it is helpful to study the response of a
2D hole systems to an external magnetic field $B$. For zero magnetic
field, the system is characterized by a large quadrupole moment
close to its maximum value. This result reflects the fact that the
HH-LH splitting is the most important effect in a 2D HH system and
the multipole expansion (\ref{eq:rhodecomp}) is the natural language
to describe this effect. A perpendicular magnetic field results in a
large dipole moment (i.e., a large spin polarization), as expected
from Fig.~\ref{fig:hspin}. Of course, for a perpendicular magnetic
field we also need to take into account the formation of Landau
levels. A low-density 2D electron system can become fully
spin-polarized without forming Landau levels when an in-plane
magnetic field is applied \cite{oka99, tut03}. When the Zeeman
energy becomes larger then the ($B$ dependent) Fermi energy, the
minority spin subband is completely depopulated and the system
becomes fully spin-polarized. In 2D HH systems, on the other hand,
the spin polarization is generally suppressed when an in-plane
magnetic field is applied \cite{win04a, win04b}. As an example,
Fig.~\ref{fig:mumo} shows the calculated response of a 2D HH system
in a symmetric (100) GaAs-Al$_{0.3}$Ga$_{0.7}$As quantum well with
hole density $p = 5 \times 10^{10}$~cm$^{-2}$ and well width
$w=150$~{\AA}.

We can see in Fig.~\ref{fig:mumo} that the complete depopulation of
the HH minority spin subband due to $B_\|$ does not imply full spin
polarization of the system \cite{win04a, win04b}. In
Fig.~\ref{fig:mumo}, the minority spin subband is completely
depopulated at $B_D \approx 23.9$~T while $\tilde{\mathcal{S}} (B_D)
\approx 0.144$. Here we use the tilde to indicate that we have
normalized $\rho_1 = \mathcal{S}$ with respect to the total 2D
density $2\rho_0$. We see that the spin polarization
$\tilde{\mathcal{S}} (B_\|)$ is always much smaller than
$3/2\sqrt{5} \approx 0.67$, the value of $\tilde{\mathcal{S}}$ in a
fully spin-polarized HH system and it is unrelated with the
depopulation of the minority spin subband [Fig.~\ref{fig:mumo}(a)].
Most surprisingly we even have a sign reversal of the spin
polarization vector at $B_\| \approx 18.4$~T which is a unique
feature of 2D HH systems \cite{win04a}. The derivative of
$\tilde{\mathcal{S}} (B_\|)$ is discontinuous at $B_\| = B_D$. At
$B_\| = 0$ the quadrupole moment $\tilde{\mathcal{Q}}$ is slightly
smaller than $1/2$, the value of $\tilde{\mathcal{Q}}$ in a pure HH
system. This is a consequence of the $\kk_\|$-induced HH-LH mixing
which was fully taken into account in Fig.~\ref{fig:mumo}. For $B_\|
> 0$ we observe only a small decrease of $\tilde{\mathcal{Q}}$. This
is due to the fact that the HH-LH splitting $E^h_1 - E^l_1 \approx
6.7$~meV [i.e., the denominator in Eq.\ (\ref{eq:zeeman_par})] is
the largest energy scale in the system so that the HH states have a
frozen angular momentum perpendicular to the 2D plane, as
illustrated in Fig.~\ref{fig:hspin}. For comparison, we note that
the Zeeman energy splitting of the HH subband at $B_D$ is $\sim
0.4$~meV.

It is remarkable that the octupole moment $\tilde{\mathcal{O}}$ at
$B_\| = B_D$ is close to $1/\sqrt{2} \approx 0.71$, the largest
possible value of $\tilde{\rho}_3$ in a 2D HH system. This value
is essentially independent of whether we use the Luttinger
Hamiltonian (\ref{eq:lutt_spher}) without or with the octupole term
proportional to $\zeta$. This result reflects the fact that, unlike
the simpler case of spin-1/2 electron systems, we \emph{cannot}
establish a simple one-to-one correspondence between the multipoles
in the Hamiltonian (\ref{eq:lutt_spher}) and the multipoles of the
same degree in the spin density matrix (\ref{eq:rhodecomp}). On the
other hand, these findings suggest that an in-plane magnetic field
$B_\|$ provides an efficient tool to study 2D HH systems with a
large octupole moment but with a small dipole moment (i.e., with a
small spin polarization). Note that we can use an \emph{in-plane}
magnetic field to obtain a significant \emph{out-of-plane} spin
polarization using the anisotropic Zeeman splitting (the component
$g_{xz}^\ast$ of the $g$ factor) on low-symmetry surfaces such as
the (113) surface, see Fig.~\ref{fig:gfak_anis_gaas}
(Ref.~\cite{win05}).

\section{Spin Precession of Holes}
\label{sec:hprec}
\subsection{General Analysis}

Before discussing hole spin precession, we will review briefly
Larmor precession in electron systems \cite{lanIV1e}. It is well
known that the dipole moment (spin polarization) of a spin-$1/2$
system performs a precessing motion in an external magnetic field
$\vek{B}$. It can be derived via the Heisenberg equation of motion
(HEM) for the spin operator $\vek{M}_1 = \vek{\sigma}$ propagating
due to a Zeeman term $\mathcal{H}_1 \equiv \frac{1}{2}g^\ast
\bohrmag \; \vek{\sigma} \cdot \vek{B}$,
\begin{equation}
  \label{eq:spinprec}
  \td{\vek{\sigma}}{t}
  = \frac{1}{i\hbar} \; [\vek{\sigma}, \mathcal{H}_1]
  = g^\ast \bohrmag \; \vek{\sigma} \times \vek{B} \,,
\end{equation}
where $\vek{\sigma}$ at time $t=0$ is the vector of Pauli spin
matrices. If the only spin-dependent term of the full Hamiltonian is
$\mathcal{H}_1$, the HEM (\ref{eq:spinprec}) is valid also for
particles with $j > 1/2$ (with $\vek{\sigma}$ replaced by the
appropriate angular momentum matrices for $j > 1/2$). Taking the
expectation value of Eq.\ (\ref{eq:spinprec}) with respect to a
state $|\psi\rangle$ yields $\langle \dot{\vek{\sigma}} \rangle =
g^\ast \bohrmag \; \langle\vek{\sigma}\rangle \times \vek{B}$, which
can be interpreted as Ehrenfest's theorem \cite{sak94} applied to a
spin-$j$ system.

In the context of conventional spin precession \cite{lanIV1e} it
appears natural that the right-hand side of Eq.\ (\ref{eq:spinprec})
can be expressed as a linear combination of spin operators
$\sigma_\fm$. This implies that the HEM for the components $\sigma_\fm$
of $\vek{\sigma}$ are closed. Using the language of multipole
moments, we see here that for spin-$1/2$ systems the HEM of the
dipole $\vek{M}_1 = \vek{\sigma}$ is decoupled from the HEM of the
monopole $M_0$. Obviously, the HEM for $M_0$ is trivial, $\dot{M}_0
= 0$, which reflects the conservation of the probability density. We
also note that Eq.\ (\ref{eq:spinprec}) preserves the length of the
Bloch vector $\langle \vek{\sigma} \rangle$, i.e., $d |\langle
\vek{\sigma} \rangle| / dt = 0$. This is equivalent to the statement
that the magnitude of the dipole moment $\vek{\rho}_1 = \vekc{S}$
does not depend on time, $\mathcal{S} (t) = \mathrm{constant}$,
which reflects the conservation of energy in the system. (Here we
ignore scattering and spin relaxation \cite{cul07}.)

We proceed to study the corresponding equations of motion of the
multipoles $\vek{M}_\fj$ of spin-$3/2$ systems. While the dipole term
in the multipole expansion of the Hamiltonian can always be
interpreted as a Zeeman-like term with an external or effective
magnetic field, no such interpretation is possible for the higher
multipoles in Eq.\ (\ref{eq:lutt_spher}) including the quadrupole
term. Therefore, the simple picture of a spin precessing around an
effective Zeeman field is not applicable to holes. Nevertheless, the
spin dynamics of hole systems {\it can} be viewed as a precession,
if precession is understood as a nontrivial {\it periodic} motion in
spin space described by an equation of the type $\td{\mathcal{S}}{t}
= \frac{i}{\hbar} [\mathcal{H}, \mathcal{S}]$ for a suitably
generalized spin operator $\mathcal{S}$ and spin Hamiltonian
$\mathcal{H}$. However, the HEM for the different $\vek{M}_\fj$ cannot
be decoupled which has important consequences for the spin
precession of hole systems \cite{cul06}. 

Equation (\ref{eq:lutt_spherg}) suggests that we study first the HEM
\begin{equation}
  \label{eq:momentprec}
  \td{\vek{M}_\fj}{t} =
  \frac{1}{i\hbar} \; [\vek{M}_\fj, \mathcal{H}_{\fj'}] \, .
\end{equation}
Making use of Eq.\ (\ref{eq:lutt_spherg}), this equation can be
decomposed into
\begin{equation}
\td{\vek{M}_{\fj}}{t} = \frac{a_{\fj'}}{i\hbar}  \,
[\vek {M}_{\fj}, \vek{M}_{\fj'}]\cdot\vekc{K}_{\fj'} ,
\end{equation}
where the dot product is between $\vek{M}_{\fj'}$ and
$\vekc{K}_{\fj'}$. The commutator is an antisymmetric tensor product
that may be further decomposed into multipoles using standard
angular momentum algebra \cite{edm60}:
\begin{equation}\label{eq:commutator}
[M_{\fj \fm}, M_{\fj'\fm'}] = \sum_{\mathfrak{J},\mathfrak{M}}
C_{\fj \fm\, \fj'\fm'}^{\mathfrak{J} \mathfrak{M}}
\left[1 - (-1)^{(\fj + \fj' + J)} \right] M_{\mathfrak{J} \mathfrak{M}}.
\end{equation}
Here, the $C_{\fj \fm \, \fj'\fm'}^{\mathfrak{J} \mathfrak{M}}$ are
Clebsch-Gordan coefficients for which the phase convention of Ref.\
\cite{edm60} has been adopted. Angular momentum conservation
constrains the sum to $|\fj-\fj'| \le {\mathfrak{J}} \le \fj + \fj'$
and $-{\mathfrak{J}} \le {\mathfrak{M}} \le {\mathfrak{J}} $. The
multipole components $M_{\mathfrak{J} \mathfrak{M}}$, which appear
on the RHS of Eq.\ (\ref{eq:commutator}), are those satisfying the
condition $\fj + \fj' + {\mathfrak{J}} = \mathrm{odd}$. The invariant
decomposition of the RHS of Eq.\ (\ref{eq:momentprec}) is summarized
in Table~\ref{tab:asymirrdec} \cite{asymprod}. As discussed above, a
dipole $\vek{M}_1$ evolving under the action of $\mathcal{H}_1$
describes the well-known Larmor precession of holes \cite{lanIV1e}.
Table~\ref{tab:asymirrdec} shows, as expected, that in this case the
RHS of Eq.\ (\ref{eq:momentprec}) yields only $\vek{M}_1$, i.e., the
HEM for a dipole, which couples to a magnetic field through a Zeeman
term, is closed. However, the most remarkable entry in the table is
the one for $\vek{M}_2$ propagating in time due to a quadrupole
$\mathcal{H}_2$, the spin-orbit interaction that gives rise to the
HH--LH coupling. We see here that the HEM for the components of
$\vek{M}_2$ are not closed. A quadrupole $\vek{M}_2$ precessing in a
quadrupole field ``decays'' into a dipole $\vek{M}_1$ and an
octupole $\vek{M}_3$. This implies that spin precession of an
initially unpolarized system can give rise to spin polarization,
even though $B=0$. [As mentioned above, we use the term spin
precession for any HEM (\ref{eq:momentprec}) with $\fj, \fj' > 0$.]

\begin{table}[tbp]
  \caption{\label{tab:asymirrdec}Irreducible representations
  $\mathcal{D}_{\fj}$ of $SU(2)$ of the (linear combinations of)
  multipole $\vek{M}_{\fj}$ contained in an invariant
  decomposition of $(1/i\hbar) \; [\vek{M}_\fj,
  \mathcal{H}_{\fj'}]$.}
  \centerline{$\arraycolsep 1em
   \begin{array}{c@{\hspace{2em}}cccc} \hline\hline
    & \mathcal{H}_0 & \mathcal{H}_1 & \mathcal{H}_2 & \mathcal{H}_3
    \\ \hline
         M_0  & 0 & 0 & 0 & 0 \\
    \vek{M}_1 & 0 & \mathcal{D}_1 & \mathcal{D}_2 & \mathcal{D}_3 \\
    \vek{M}_2 & 0 & \mathcal{D}_2 & \mathcal{D}_1 \oplus \mathcal{D}_3
                  & \mathcal{D}_2 \\
    \vek{M}_3 & 0 & \mathcal{D}_3 & \mathcal{D}_2
                  & \mathcal{D}_1 \oplus \mathcal{D}_3 \\ \hline\hline
  \end{array}$}
\end{table}

We want to determine and interpret the explicit time evolution of
$\rho$. This calculation is most easily carried out in the
Schr\"odinger picture, which reflects the equivalence of the
Heisenberg and Schr\"odinger pictures for this problem. Our final
results below are independent of representation. In the absence of
external fields and disorder, the density matrix satisfies the
quantum Liouville equation
\begin{equation}
\pd{\rho}{t} = \frac{i}{\hbar}\,[\rho, \mathcal{H}].
\end{equation}
Note the sign difference between the Liouville and Heisenberg
equations. The formal solution is $\rho (t) = e^{-i\mathcal{H}
t/\hbar} \rho (0) e^{i\mathcal{H} t/\hbar}$, where $e^{i\mathcal{H}
t/\hbar}$ is the time evolution operator (which can often be
evaluated in closed form).

\subsection{Precession of a Single Spin}

We consider first an example where we assume the hole spin to be
oriented initially along the $z$-direction, $m_j = + 3/2$, so that
\begin{equation}
  \label{eq:densini}
  \tilde{\rho} (t=0) = \frac{1}{2} M_{0,0}  + \frac{3}{2\sqrt{5}} M_{1,0}
  + \frac{1}{2} M_{2,0} + \frac{1}{2\sqrt{5}} M_{3,0},
\end{equation}
where the tilde indicates that $\rho$ has been normalized with
respect to the total density $2\rho_0$. This equation demonstrates
that, in general, the density matrix of holes cannot be written
simply as the sum of a monopole and a dipole. The higher multipoles
will be present, too \cite{win05}.

\begin{figure}[tbp]
 \centerline{\includegraphics[width=35mm]{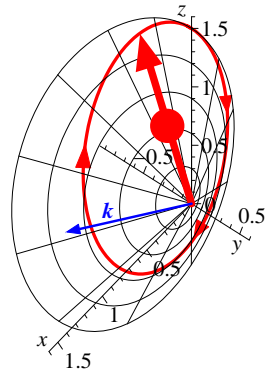}}
 \caption{\label{fig:bloch}Precession of the Bloch vector $\vek{S}$
 (bold arrow) about the wave vector $\kk$ when the angle between the
 two is initially $60^\circ$. We have assumed $\vek{S}(t=0)$ is
 parallel to the $z$ axis, and $\kk$ lies in the $xz$ plane. Taken
 from Ref.\ \cite{cul06}.}
\end{figure}

We want to restrict ourselves to $\mathcal{H} = \mathcal{H}_0 +
\mathcal{H}_2$, i.e., $B=0$. Table~\ref{tab:asymirrdec} shows that
$\rho$ evolves as a combination of a dipole, a quadrupole, and an
octupole. The implications of this fact are best seen by following
the motion of the Bloch vector $\vek{S}(t) =
\sqrt{5} \, \tilde{\vekc{S}} (t)$
\begin{equation}
  \label{eq:bloch_hh}  \arraycolsep 0.1ex
  \begin{array}[b]{rl}
\vek{S} (t) = \frack{3}{2}\big\{ & \hat{\vek{S}}_0
\left[ \cos^2 (\omega t)
    + c^4 \sin^2 (\omega t) \right] \\[1.0ex]
  & \displaystyle
  {} + \hat{\vek{\kappa}}\, s\,c\, (1 + c^2)\, \sin^2 (\omega t)
 \\[1.2ex]
  & \displaystyle
  {} + (\hat{\vek{S}}_0 \times \hat{\vek{\kappa}}) \, 2\, s\, c\,
\sin(\omega t) \cos(\omega t) \big\},
  \end{array}
\end{equation}
where the hat denotes unit vectors; we have $\vek{S}_0 = \vek{S}
(t=0)$; $c = \hat{\vek{S}}_0 \cdot \hat{\kk}$ is the cosine of the
angle between $\hat{\vek{S}}_0$ and $\kk$; $s$ is the sine of the
same angle; and $\vek{\kappa} = \kk - (\hat{\vek{S}}_0 \cdot \kk)
\hat{\vek{S}}_0$ is the vector orthogonal to $\vek{S}_0$ in the
$(\vek{S}_0, \kk)$ plane. We see that the trajectory $\vek{S} (t)$
in spin space is independent of the Luttinger parameters; only the
frequency $\omega = (E_h - E_l)/2\hbar \simeq \bar{\gamma}\hbar
k^2/m_0 $ depends on $\bar{\gamma}$ and $k$. When $\vek{S}_0$ is
parallel to $\kk$ (i.e., $c = 1$), we get $\vek{S} (t) = \vek{S}_0$,
which is due to the fact that the initial state is an eigenstate of
the Hamiltonian. In general, neither the magnitude nor the
orientation of the Bloch vector are conserved. This is illustrated
in Fig.~\ref{fig:bloch} showing $\vek{S} (t)$ for an angle of
$60^\circ$ between $\hat{\vek{S}}_0$ and $\kk$. Helicity
$\vek{S}\cdot{\kk}$ is conserved, a well-known fact about this
model, which sheds additional light on spin precession in hole
systems. Since $\td{}{t}(\vek{S}\cdot{\kk}) = 0$ and the wave vector
is not changing, $\td{\vek{S} }{t}\cdot {\kk} = 0$. Therefore,
whenever the magnitude of the spin changes, the angle between spin
and wave vector must change in order to preserve the projection of
$\vek{S}$ onto ${\kk}$. As a consequence, no nontrivial spin
precession occurs when $ \vek{S} \perp \kk $ (or $ \vek{S} \parallel
\kk$). We note that energy is also conserved for spin precession in
hole systems. Yet for $B>0$, when $\mathcal{H} = \mathcal{H}_0 +
\mathcal{H}_1 + \mathcal{H}_2$, it can be shown that, in the general
case, energy is transferred back and forth between $\mathcal{H}_1$
and $\mathcal{H}_2$ as time progresses. For an LH spin ($m_j = +1/2$)
we obtain similar to Eq.\ (\ref{eq:bloch_hh})
\begin{equation}
  \label{eq:bloch_lh}  \arraycolsep 0.1ex
  \begin{array}[b]{rl}
\vek{S} (t) =  & \hat{\vek{S}}_0
\left[ \frack{1}{2} \, +  \frack{3}{2} \, s^2 (2-3s^2) \,\sin^2 (\omega t) \right] \\[1.0ex]
  & \displaystyle
  {} + \frack{3}{2} \, \hat{\vek{\kappa}}\, s\,c\, (1 -3 c^2)\, \sin^2 (\omega t)
 \\[1.2ex]
  & \displaystyle
  {} - 3\, (\hat{\vek{S}}_0 \times \hat{\vek{\kappa}}) \, s\, c\,
\sin(\omega t) \cos(\omega t) .
  \end{array}
\end{equation}

The analysis above highlights a major advantage of our method.
Although most of the results presented here can be derived also
using wave functions, the decomposition of $\rho$ and $\mathcal{H}$
into multipoles makes the symmetry of the problem transparent
\cite{lip70}. In particular, the interdependence of the multipoles
would not be evident if one used wave functions. Moreover, the
multipoles do not rely on a particular choice of basis functions
\cite{win04b} --- in that sense the multipoles are gauge-invariant.

\subsection{Holes Turning a Corner at $B=0$: Alternating Spin
Polarization}

\begin{figure}[tbp]
 \centerline{\includegraphics[width=80mm]{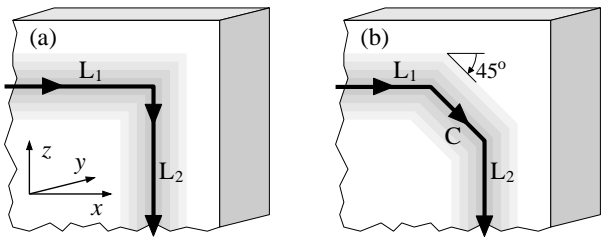}}
 \caption{\label{fig:grayson}%
 A bent structure allows holes to be driven around a corner
 \cite{gra04, gra05}.
(a) An idealized representation of the corner device.
(b) A more realistic model of the corner, as discussed in the text.
In regions C and L$_2$ the spin polarization oscillates as a
function of time. Taken from Ref.\ \cite{cul06}.}
\end{figure}

We would like to relate our work to recent experiments by Grayson
\emph{et al.} \cite{gra04, gra05} which demonstrated that a
high-quality bent heterostructure can be grown on top of a
pre-cleaved corner substrate that allows one to drive the charge
carriers around an atomically sharp $90^\circ$ corner. Grayson's
experiments were performed on a 2D electron system in a GaAs/AlGaAs
quantum well. We show here that a similar system containing holes
gives rise to fascinating new physics \cite{cul06}. The setup is
sketched in Fig.~\ref{fig:grayson}(a). We assume $B=0$ so that the
Hamiltonian is $\mathcal{H} = \mathcal{H}_0 + \mathcal{H}_2$. An
unpolarized HH wave packet travels in the 2D channel L$_1$ in the
$+x$ direction. For simplicity we assume that all orbital degrees of
freedom are integrated over, so that ${\rho}$ depends only on the
spin indices and does not depend on the wave vector $\kk$.
Nevertheless, the coupling of spin and $\kk$ is present in the
Hamiltonian through the spin-orbit interaction. The magnitudes of
the normalized moments at $t \le0$ (i.e., before reaching the
corner) are $\tilde{\mathcal{S}} = \tilde{\mathcal{O}} = 0$ and
$\tilde{\mathcal{Q}} = 1/2$. The spin quantization axis of the HH
states in L$_1$ is parallel to the $z$-direction. After the wave
packet has passed the corner, the HH states are not eigenstates of
$\mathcal{H}_2$, their spin quantization axis being perpendicular to
the spin quantization axis supported by the quadrupole field in
L$_2$. Therefore, the quadrupole and octupole moments in L$_2$
oscillate in time
\begin{subequations}
  \label{eq:m90corner}
  \begin{eqnarray}
    |\tilde{\vekc{Q}} (t)|^2 & = & (1/16) + (3/16) \cos^2 (\omega_z t) \\
    |\tilde{\vekc{O}} (t)|^2 & = & (3/16) \sin^2 (\omega_z t),
  \end{eqnarray}
\end{subequations}
with precession frequency $\omega_z = (E_h - E_l)/2\hbar \simeq
2\bar{\gamma}\hbar \pi^2/m_0 w^2$. This frequency can be tailored by
varying the width $w$ of the 2D channel. For GaAs in the spherical
approximation we have $\bar{\gamma} = 2.58$ and we take $w = 10$~nm,
yielding a precession period $2\pi/\omega_z \approx 11$~ps.

For the simplified geometry of Fig.~\ref{fig:grayson}(a) the spin
polarization $\tilde{\mathcal{S}}$ in L$_2$ remains zero, as
required by the conservation of helicity discussed above. However,
assuming a sharp $90^\circ$ corner for an \emph{ideal} 2D system is
certainly an oversimplification, even when the corresponding
\emph{quasi} 2D system has atomically sharp interfaces \cite{gra04}.
A more realistic treatment can be obtained by modeling the
transition region between the channels L$_1$ and L$_2$ as a sequence
of two $45^\circ$ corners as sketched in Fig.~\ref{fig:grayson}(b).
Once again, an unpolarized HH wave packet travels in channel L$_1$
in the $+x$ direction, the initial conditions being the same as in
the previous example. If the HH wave packet enters the central
region C at $t=0$, we obtain for the squared normalized moments in
this region
\begin{subequations}
  \label{eq:m45corner}
  \begin{eqnarray}
    \label{eq:m45corner:s}
    |\tilde{\vekc{S}} (t) |^2 & = & (9/80) \; \sin^2 (\omega_z t)\\
    |\tilde{\vekc{Q}} (t) |^2 & = & 1/64 + (15/64) \cos^2 (\omega_z t) \\
    |\tilde{\vekc{O}} (t) |^2 & = & (39/320) \;  \sin^2 (\omega_z t).
  \end{eqnarray}
\end{subequations}
Equation (\ref{eq:m45corner:s}) shows that the initially unpolarized
hole current acquires an alternating spin polarization
$\tilde{\vekc{S}} (t)$ due to spin precession at $B=0$. In Cartesian
coordinates, the Bloch vector in region C reads $\tilde{\vek{S}}(t)
= [0, -\frac{3}{4\sqrt{5}}\, \sin (\omega_z t), 0]$. When the HH
wave packet enters the channel L$_2$ it continues to precess. We get
for the spin polarization
\begin{equation}
  \label{eq:m2x45corner}
  \tilde{\mathcal{S}_y} = \frac{-3}{8\sqrt{5}}
  \left[\cos^2 \left(\frac{\omega_z T}{2} \right)\sin (\omega_z t)
  + 2\sin (\omega_zT) \sin^2 \left( \frac{\omega_zt}{2} \right)\right].
\end{equation}
Here $T$ is the time required to traverse~C which depends on the
length of C and the magnitude of the in-plane wave vector. It
determines the fraction of the spin polarization in L$_2$ that
will be oscillating. If we take the length of C to be of the order
of the channel width, namely $w=10$~nm, and the initial wave
vector $k_F = 0.1$~nm$^{-1}$, we get $T\sim 0.5$~ps. The amplitude
of $\tilde{\mathcal{S}_y}$ in this case is approximately~0.1. We
omit here the qualitatively similar but more complicated
expressions for $\tilde{\vekc{Q}}(t)$ and $\tilde{\vekc{O}}(t)$.
We note also that the approach in Fig.~\ref{fig:grayson}(b) can be
further extended in a transfer-matrix-like approach in order to
describe more complicated geometries.

\section{Spin Relaxation in Hole Systems}
\label{sec:relax}

Spin relaxation in spin-1/2 electron systems has received
considerable attention \cite{dya72, dya86, pik84, dya71, dya06,
kis00, son02, hua03, ohn99, sli90, dzh04, ave99, ave02, ave06,
gri02, ell54, yaf63, qi03, mis04, lau01, kai04, bro04, ble04,
ohn07}. For electrons the spin-orbit interaction can always be
represented by a Zeeman-like term (\ref{eq:rash_effB}) with an
effective wave vector-dependent magnetic field $\vekc{B} (\kk)
\equiv \hbar\, \vek{\Omega} (\kk)$. The electron spin precesses
about this field with frequency $\Omega = |\vek{\Omega} (\kk)|$. An
important parameter is the product of the frequency $\Omega$ times
the momentum relaxation time $\tau_p$. In the ballistic (clean)
regime no scattering occurs, so that $\Omega\tau_p \rightarrow
\infty$. The weak scattering regime is characterized by fast spin
precession and little momentum scattering, yielding $\Omega \,
\tau_p \gg 1$. In the strong momentum scattering regime $\Omega \,
\tau_p \ll 1$. Electron systems are often in the strong scattering
regime and most past work has concentrated on this case.

As discussed in Sec.~\ref{sec:muexp}, for spin-3/2 holes the
spin-orbit interaction cannot be written as an effective field, and
spin precession is qualitatively different (Sec.~\ref{sec:hprec}).
Since spin-orbit coupling is more important in the valence band,
hole spin information is lost faster, and the relative strengths of
spin-orbit coupling and momentum scattering can vary. Yet spin
relaxation of spin-3/2 holes has also been studied to a lesser
extent, both experimentally \cite{hil02} and theoretically
\cite{ave02, dya84a, ser05, lue06, uen90, fer91}.

Recently we were able to derive a general unifying quantitative
theory for the return to equilibrium of excess spin polarizations in
the conduction and valence bands of semiconductors brought about by
the interplay of spin precession and momentum scattering
\cite{cul07}. Spin polarization decay in different regimes of
momentum scattering in spin-1/2 electron and spin-3/2 hole systems
contains considerable rich and novel physics. For example, spin
polarization decay has often been assumed to be proportional to
$e^{-t/\tau_s}$, where $\tau_s$ is referred to as the \emph{spin
relaxation time}. However, if the magnitude of the spin-orbit
interaction is anisotropic (as is usually the case in systems
studied experimentally), spin-polarization decay can occur even in
the absence of momentum scattering. This process is characterized by
a non-exponential decay and is sensitive to the initial conditions,
and cannot therefore be described by a spin relaxation time. Weak
momentum scattering introduces a spin relaxation time $\tau_s
\propto \tau_p$ (unlike strong momentum scattering, which gives the
well-known~\cite{dya72, pik84} trend $\tau_s \propto \tau_p^{-1}$),
yet even in the presence of weak momentum scattering a fraction of
the polarization may survive at long times. Indeed, in the ballistic
and weak momentum scattering regimes, the concept of a spin
relaxation time is of very limited applicability and in general does
not provide an accurate description of the physics of spin
polarization decay.

\subsection{Time Evolution of the Density Matrix}

We assume a nonequilibrium spin polarization has been generated in a
homogeneous, unstructured system and study its time evolution in the
absence of external fields. The system is described by a density
matrix, which in principle has matrix elements diagonal and
off-diagonal in momentum space. Since the spin operator is diagonal
in the wave vector $\kk$, we will only be concerned with the part of
the density matrix diagonal in momentum space, which is denoted by
$\rho_\kk$.

The spin density is given by $\langle \vek{S} \rangle \equiv \trace
(\vek{S} \rho_\kk) = \trace (\vek{S} \bar{\rho}_k)$, where $\vek{S}$
is the spin operator, the trace includes $\kk$ and spin, and the
overline represents averaging over directions in momentum space.
Only the isotropic part $\bar{\rho}_k$ of the density matrix is
responsible for spin population decay \cite{pik84}. It is therefore
convenient to divide $\rho_\kk$ into $\rho_\kk = \bar{\rho}_k +
g_\kk$, where $g_\kk$ is the anisotropic part of $\rho_\kk$. Based
on the quantum Liouville equation, we obtain an equation describing
the time evolution of $\rho_\kk$ (Ref.\ \cite{ave02}), which in turn
is split into a set of equations for $\bar{\rho}_k$ and $g_\kk$
similar to those found by Pikus and Titkov \cite{pik84}:
\begin{subequations}
\begin{eqnarray}\label{eq:rhobarga}
\pd{\bar{\rho}_k}{t} + \frac{i}{\hbar} \, \overline{[\mathcal{H}, g_\kk]}
& = & 0,
\\ [1ex]\label{eq:rhobargb}
\pd{g_\kk}{t} + \frac{i}{\hbar} \, [\mathcal{H}, g_\kk]
+ \frac{g_\kk}{\tau_p} & = & - \pd{\bar{\rho}_k}{t}
- \frac{i}{\hbar} \, [\mathcal{H}, \bar{\rho}_k] .
\hspace{2em}
\end{eqnarray}
\end{subequations}
Here, $\mathcal{H}$ is, in general, the full Hamiltonian; yet the
spin-diagonal part of $\mathcal{H}$ commutes with $\rho_\kk$ so that
it is not relevant for spin relaxation. We assume elastic,
spin-independent scattering, implying that the collision term
involving $\bar{\rho}_k$ vanishes \cite{pik84}. For isotropic ($s$
wave) scattering due to, e.g., screened impurities the remainder is
proportional to the inverse of the scalar momentum relaxation time
$1/\tau_p$ (Ref.~\cite{multi_tau}). A solution to Eq.\
(\ref{eq:rhobargb}) can be obtained by making the transformation
$g_\kk = e^{- i \mathcal{H} t/\hbar} g_{\kk}^H \, e^{i \mathcal{H}
t/\hbar}$, which is analogous to the customary switch to the
Heisenberg picture. Substituting this solution into Eq.\
(\ref{eq:rhobarga}) yields \cite{cul07}
\begin{equation}
  \label{eq:rhobarJ}
  \arraycolsep 0.1ex \begin{array}[b]{rl} \displaystyle
    \pd{\bar{\rho}_k}{t} + \frac{i}{\hbar\tau_p} \int_0^{t} \! dt'
  & \displaystyle
    e^{- (t - t')/\tau_p} \, \overline{e^{- i \mathcal{H} (t - t')/ \hbar} \,
    [\mathcal{H}, \bar{\rho}_k(t') ] \,
    e^{i \mathcal{H} (t - t') / \hbar}}
  \\ [2.2ex] & \displaystyle
    = -  \frac{i}{\hbar} \, e^{-t/\tau_p}\,
    \overline{e^{-i \mathcal{H} t / \hbar} \,
    [\mathcal{H}, \rho_{\kk 0}]\, e^{i \mathcal{H} t / \hbar}} ,
  \end{array}
\end{equation}
where $\rho_{\kk 0}$ is the initial value $\rho_\kk (t=0)$. This
equation describes the precession-induced decay of spin polarization
in all regimes of momentum scattering for both spin-1/2 electrons
and spin-3/2 holes in semiconductors. It does not anticipate any
particular form of spin polarization decay, such as exponential
decay.

The form of the initial density matrix $\rho_{\kk 0}$ is important
and lies at the root of the physics discussed in the remainder of
this section. In general $\rho_{\kk 0}$ has two contributions,
$\rho_{\kk 0} = \rho_{\kk 0}^\| + \rho_{\kk 0}^\perp$. The component
$\rho_{\kk 0}^\|$ commutes with $\mathcal{H}$ and $\rho_{\kk
0}^\perp$ is simply the remainder. $\rho_{\kk 0}^\|$ is a matrix
that is parallel to the Hamiltonian, and represents the fraction of
the initial spin polarization that does not precess, or
alternatively the fraction of the initial spins that are in
eigenstates of the Hamiltonian. $\rho_{\kk 0}^\perp$ is orthogonal
to the Hamiltonian, and represents the fraction of the initial spin
polarization that does precess.

\subsection{Regimes of Spin Polarization Decay in Spin-3/2 Hole
Systems}

In general, Eq.\ (\ref{eq:rhobarJ}) allows one to distinguish and
describe several different regimes of spin polarization decay by
comparing the precession frequency $\Omega (\kk)$ with the momentum
relaxation time $\tau_p$ \cite{cul07}. In the following we consider
spin-3/2 holes described by the Luttinger Hamiltonian
(\ref{eq:lutt_lutt}). We work first in the spherical approximation,
Eq.\ (\ref{eq:lutt_spher}).

\subsubsection{Exponential Decay in the Strong Momentum Scattering Regime}

A solution to Eq.\ (\ref{eq:rhobarJ}) characterizing
\emph{relaxation} is understood as \emph{exponential} decay of the
form
\begin{equation}
  \bar{\rho}_k (t) = e^{-\Gamma_s t}\,\bar{\rho}_{k 0},
\end{equation}
where $\Gamma_s$ is generally a second-rank tensor that represents
the inverse of the spin relaxation time $\tau_s$. Such a simple
solution of Eq.\ (\ref{eq:rhobarJ}) does \emph{not} exist in
general, but for strong momentum scattering ($\Omega \tau_p \ll 1$)
the RHS of Eq.\ (\ref{eq:rhobarJ}) can be neglected. Then
substituting for $\bar{\rho}_k$ and $\mathcal{H}$ in Eq.\
(\ref{eq:rhobarJ}) yields an exponential decay of the spin
population with a relaxation time $\tau_s \propto \tau_p^{-1}$. This
trend is well-known for Dyakonov-Perel spin relaxation in electron
systems \cite{dya72, pik84}.

For spin-3/2 holes we get $\Gamma_s = \tau_s^{-1} \, \openone$,
showing that (for a given wave vector) the relaxation times for all
spin components are equal,
\begin{equation}
\frac{1}{\tau_s} = {\textstyle\frac{2}{5}} \, \Omega^2\, \tau_p
= \frac{8}{5} \left(\frac{\hbar \bar{\gamma} k^2}{m_0}\right)^2 \tau_p,
\end{equation}
where $\Omega (\kk) = (E_\mathrm{LH} - E_\mathrm{HH}) / \hbar =
2\hbar\bar{\gamma} k^2/m_0$. The spin relaxation times $\tau_s$ for
HH states become effectively smaller than $\tau_s$ for LH states if
we take into account that HH states are normally characterized by a
larger Fermi wave vector $k_F$. Despite the qualitatively different
spin precession, the situation is overall rather similar to electron
spin relaxation and can be explained in terms of the same random
walk picture familiar from the study of electron spin relaxation
\cite{dya72, pik84}.

\subsubsection{Ballistic Regime in the Spherical Approximation}

In the opposite limit $\tau_p \rightarrow \infty$ the second term on
the LHS of Eq.\ (\ref{eq:rhobarJ}) can be neglected. Then Eq.\
(\ref{eq:rhobarJ}) is solved by
\begin{equation}
  \label{eq:ballistic}
  \bar{\rho}_k (t) = \bar{\rho}_{\kk 0}^\| +
  \overline{e^{-i \mathcal{H} t/\hbar} \, \rho_{\kk 0}^\perp \, e^{i
  \mathcal{H} t/\hbar}}
\end{equation}
which describes a spin precession of the initial spin
polarization $\rho_{\kk 0}$. In the spherical approximation
(\ref{eq:lutt_spher}) for the Luttinger Hamiltonian, an initial spin
polarization will oscillate indefinitely since $\Omega$ is the same
for all holes on the Fermi surface.

\subsubsection{Weak Momentum Scattering Regime in the Spherical
Approximation}

In the regime of weak momentum scattering the solution to Eq.\
(\ref{eq:rhobarJ}) may be written approximately as
\begin{equation}\label{eq:rhobarweak}
\bar{\rho}_k (t) = \bar{\rho}_{\kk 0}^\| + e^{-t/\tau_p} \, \overline{e^{-i
\mathcal{H} t'/\hbar} \, \rho_{\kk 0}^\perp \, e^{i \mathcal{H} t'/\hbar}},
\end{equation}
Since the momentum scattering rate $1/\tau_p$ is small, the term
under the overline is taken to lowest order in $1/\tau_p$. The
second term on the RHS of Eq.\ (\ref{eq:rhobarweak}) describes
damped oscillations with amplitude decaying exponentially on a scale
$\propto \tau_p$. The fraction $\vek{S}_{\kk 0}^\|$ of the spin
polariazarion corresponding to $\rho_{\kk 0}^\|$ survives at long
times. We can determine $\vek{S}_{\kk 0}^\|$ by averaging Eqs.\
(\ref{eq:bloch_hh}) and (\ref{eq:bloch_lh}) over time $t$ and
directions of $\kk$, showing that $S_{\kk 0}^\| = \frac{6}{10}
S_{\kk 0}$ for both HH and LH states. This result does not depend on
the Luttinger parameters or the Fermi wave vector and will therefore
be the same in any system described by the Luttinger Hamiltonian
$\mathcal{H}_s$. The remaining polarization $S_{\kk 0}^\|$ decays
via spin-flip scattering as discussed in Refs.~\cite{uen90, fer91}.

\subsubsection{Cubic-Symmetry Terms and Dephasing}

Dephasing is introduced if the term $\mathcal{H}_\mathrm{c}$ with
cubic symmetry is included in the Luttinger Hamiltonian. The
cubic-symmetry terms contained in Eq.\ (\ref{eq:luttcub}) are
usually neglected in charge and spin transport without a significant
loss of accuracy. Due to the presence of $\mathcal{H}_\mathrm{c}$,
the energy dispersion relations and therefore $|\vek{\Omega} (\kk)|$
depend on the direction of $\kk$. Spins on the Fermi surface thus
precess with incommensur<able frequencies and once they are out of
phase they never all get in phase again. Even in the ballistic limit
this process results in a non-exponential spin decay \cite{gla05z,
cul07} with a characteristic time $\tau_d \propto
\bar{\Omega}^{-1}$, referred to as the dephasing time $\tau_d$. In
general (in particular for 3D systems), an intermediate situation is
realized where the spin polarization is reduced because of
dephasing, but it remains finite. The surviving part is identified
with $\rho_{\kk 0}^\|$ in the initial density matrix. This process
is referred to as \emph{incomplete} spin dephasing.

\begin{figure}[tbp]
 \centerline{\includegraphics[width=85mm]{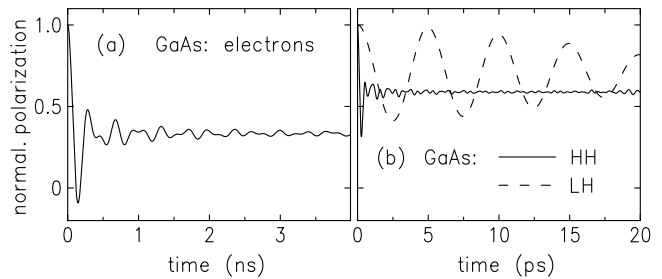}}
 \caption{\label{fig:deph} Incomplete spin dephasing of (a) electron
 spins in the $k^3$-Dresselhaus model and (b) of heavy-hole spins
 (solid line) and light-hole spins (dashed line) in bulk GaAs in
 the ballistic limit. The vertical axis shows the normalized spin
 polarization $\vek{S}(t) \cdot \hat{\vek{S}}_0 / |\vek{S}_0|$. The
 initial spin polarization $\vek{S}_0$ is assumed to point along
 [001]. In (a) the Fermi energy is $E_F = 5.7$~meV; in (b) we used
 $E_F = -1.0$~meV. Note the different time scales in (a) and (b).}
\end{figure}

Our numerical calculations exemplified in Fig.~\ref{fig:deph} show
the incomplete dephasing of electrons and holes in bulk GaAs. For
electrons, dephasing is caused by the $k^3$-Dresselhaus model
\cite{dre55a}. At long times the initial spin polarization settles
to a value $\approx 0.33$, which is independent of any system
parameters, including the spin-orbit constant. For holes, the
initial spin polarization falls to a fraction much higher than in
the electron case. It decays more slowly for the LHs, for which the
Fermi surface is nearly spherical, than for the HHs, for which the
Fermi surface deviates significantly from a sphere. Note also that
for a given Fermi energy the HH Fermi wave vector $k_F^\mathrm{HH}$
is much larger than $k_F^\mathrm{LH}$ so that the HH states precess
faster than the LH states. At long times the spin polarization
settles to a value $\approx 0.59$ for the HH states and $\approx
0.70$ for the LH states. These nonuniversal values differ slightly
from the universal value $6/10$ of the surviving spin polarization
obtained in the spherical approximation.

In Fig.~\ref{fig:deph} we assumed that the initial spin polarization
is isotropic in $\kk$ space. It is known \cite{dym76} that optically
excited spin distributions are highly anisotropic. Indeed, it turns
out that optically oriented heavy or light holes in 3D do not
precess at all (i.e., $\rho_{\kk 0}^\perp = 0$) which is similar to
2D electrons in a symmetric QW on a [110] surface \cite{dya86,
ohn99}.

\section{Conclusions}
\label{sec:conclusion}

We have reviewed spin orientation in semiconductor hole systems that
are characterized by an effective spin $j=3/2$. We showed that the
Zeeman splitting and Rashba spin splitting in hole systems are
qualitatively different from their counterparts in electron systems.
A systematic understanding of the unusual spin-dependent phenomena
in hole systems can be gained using a multipole expansion of the
spin density matrix. As an example we discussed the spin precession
in hole systems that can give rise to an alternating spin
polarization. Finally we discussed the qualitatively different
regimes of hole spin polarization decay in clean and dirty samples.

\acknowledgments

The authors appreciate stimulating discussions with C.~Lechner,
E.~P.\ De Poortere, and E.~Tutuc. Also, we are grateful to
D.~Wasserman and S.~A.\ Lyon for growing the wafers for our
experiments. We thank the DOE, ARO, NSF and the Alexander von
Humboldt Foundation for support. The research at Argonne National
Laboratory was supported by the US Department of Energy, Office of
Science, Office of Basic Energy Sciences, under Contract No.\
DE-AC02-06CH11357.


\begin{thebibliography}{10}
\bibitem{asymprod}
Table~\ref{tab:asymirrdec} is valid quite generally. It can easily be
  generalized to systems with $j > 3/2$ including $j$ integer.
\bibitem{multi_tau}
In a more elaborate approach different relaxation times characterize the
  spherical components of $\rho$ and corrections due to spin-orbit coupling are
  present in the collision term. It can be shown that neither change affects
  the physics qualitatively.
\end{thebibliography}
\end{document}